\documentclass{gen-p-l}
\usepackage{epsfig,eepic}

\theoremstyle{definition}

\theoremstyle{remark}

\numberwithin{equation}{section}

\usepackage{epsfig}
\begin{document}

\title{ On Field Theory of Open Strings, Tachyon Condensation and Closed Strings }

\author{ Samson L. Shatashvili}
\address{Department of Physics, Yale University, New Haven, CT  06520-8120}
\email{samson.shatashvili@yale.edu}
\date{April 2, 2001}
\thanks{On leave of absence from St. Petersburg Branch of Steklov Mathematical 
Institute, Fontanka, St.Petersburg, Russia. }

\begin{abstract}
I review the physical properties of different vacua
in the background independent open string field theory.
\end{abstract}

\maketitle


One of the most popular candidates for field theory of open strings is the classical action
proposed by Witten in 1985 \cite{1} - cubic {\bf CS} action. The first test that any string
field theory action shall be subject to is to recover all tree level as well as loop amplitudes
(which are independently known exactly from world-sheet approach) by standard field theory
methods, and cubic action does produce correct tree level amplitudes \cite{2}. 
It seems that if tree
level amplitudes are recovered by unitarity one can reconstruct all perturbative, loop,
amplitudes. But, it is known that one loop diagram in any field theory of open strings shall
contain the closed string pole (the cylinder diagram can be viewed as one loop in open string
theory, or equivalently as a tree level propagator for closed strings; thus there shall be a
pole for all on-shell closed string momenta); from the point of view of open string field
theory these new (closed string) poles violate unitarity since corresponding degrees of freedom
are not (at least in any obvious way) present in the particle spectrum of classical lagrangian.
This situation is similar to the one in anomalous gauge theory, but in the latter one can choose
the representation for fermions such that anomaly cancels. In open string field theory case one
can make arrangements when closed string poles decouple (for example in topological or 
non-commutative setup)
but it seems very interesting to study other possibilities.

One can think of two options: 1. find the closed string degrees of freedom as ``already being
present" in given classical open string field theory lagrangian, or 2. introduce them as
additional degrees of freedom (for example by adding corresponding string field together with
its lagrangian plus the interaction term with open string field).

In the lines of the option 2 the solution to the above problem has been found many years ago by B.
Zwiebach (see recent version \cite{3} which also contains 
references on old work). In this approach one shall take care of the problem of multiple counting
when closed string field and its lagrangian is added - the same Feynman diagram will come both from
open and closed string sectors; thus one shall make sure that each diagram is properly counted
only once; the solution of this problem is rather complicated and requires the detailed
knowledge of the string amplitudes to all loop order from world-sheet approach.\footnote{One 
shall note that adding closed string fields to open string field theory lagrangian is very similar 
to the formalism of \cite{4} for quantization of anomalous gauge theories.}

It is very interesting to explore option 1 instead. In order to do so one needs to have a
truly background independent open string 
field theory. Unfortunately such a theory has not yet been written
although the theory which doesn't depend (at least formally) on the choice of open string
background is known \cite{5}, \cite{6}, \cite{7}, \cite{8} (it also passes the test of reproducing
all tree level on-shell amplitudes in a very simple way 
since  corresponding action on-shell is given by world-sheet
partition function on the disk as it was explained in \cite{6}, \cite{8}).

One might try to make option 1 more precise by exploring the idea of closed string degrees
of freedom being some kind of classical, solitonic, solutions of open string field theory. More
concretely: start with open string field theory (background independent) and find the new
background - closed string. This seems to be a natural way of ``reversing the arrow" which
describes D-branes (open strings in case of space-filling branes) as solitons in closed string
theory \cite{9}.

The above line of thoughts suggests that we shall change the point of view about branes in
general and think about them as solitons in open string field theory rather than in closed one.
In fact, if one considers the maximal dimensional brane (or brane-anti-brane system) it is very
easy to think \cite{10} about lower-dimensional branes as solutions of classical equations 
of motion for
corresponding open string field theory action in the formalism of background independent open
string field theory. The latter is defined via {\it the action $S(t)$ - a functional on the
space of boundary conditions for bosonic string on the disc with critical points $t=t_*
\rightarrow$ being the conformal boundary conditions.} Since for the trivial bulk backgrounds
($\Delta$ operator on world-sheet) mixed Dirichlet-Neumann conditions (D-branes) are conformal
(in fact in this case these are only conformal boundary conditions) they shall correspond to
critical points of space-time action for open strings \cite{6}, \cite{10}.

Another motivation for the search of closed strings in open string field theory
(at least for the present author) comes from Matrix Strings
\cite{11}.
If we take the soliton corresponding to $ N$ $D1$-strings in open string field theory
and look on the dynamics of
collective modes we find (in strong coupling for 2d theory on $D1$ and large $N$)
the spectrum of closed strings in the same space-time
where open strings live,
thus we might ask the question whether these closed string degrees of freedom
are already present in original theory of open strings where we had $D1$'s as solitons
in the beginning. This observation also might help in  making
contact with option 1 described above. One shall mention that the search for
closed strings in open string field theory has a long history and was originated
in \cite{12}; in the above context, more in the lines of
 current developments related to
D-branes, the interest has been revived in \cite{10}.

The conjectures put forward by A. Sen \cite{13}, \cite{14} made it possible to study these
questions in much more precise terms. For simplicity one considers the open bosonic string in
26 dimensions ($D25$, or any $Dp, p < 26$) which contains tachyon and is unstable. Three
conjectures made by A. Sen are:

1. Tachyon potential takes
the form:
\begin{equation}
\begin{split}
\label{senz}
V(T) = M f(T),
\end{split}
\end{equation}
with $M$-mass of D-brane and
$f$ - universal function independent of the background where brane is
embedded. The conjecture of Sen states that $f(T)$ has a stationary
point (local minimum) at some $T=T_c < \infty$ such that
\begin{equation}
\begin{split}
\label{con}
f(T_c)=-1
\end{split}\end{equation}
and thus
$M+V(T_c) = M (1+f(T_c))=0.$

2. There are soliton configurations on unstable D-branes which correspond
to lower-dimensional branes.

3. New vacuum, at $T_c$, is a closed string vacuum and in addition there are no open string
degrees of freedom.

One might be tempting to amplify the {\bf Conjecture 3} a bit \cite{10} and
claim that in properly defined
open string field theory there shall be an expansion around new critical point which will
describe the theory of closed strings (without open string sector; of course
in this theory of closed strings we again can discuss open strings as solitonic
branes). Apriori there is no reason to assume that such expansion should exist since the
potential not necessarily shall be analytic, but one just can hope to see whole closed string
sector and not just vacuum by starting from classical lagrangian for open strings. We will
comment on this question at the end of this talk. One shall note that the picture described
below together with 
the corresponding tachyon potential is very attractive from the point of view of
applications of string theory to the phenomenology related to branes and also
to stringy cosmology.

We will address these problems in the formalism of \cite{5}, \cite{6}, \cite{7}, \cite{8}
and present the exact tachyon lagrangian up to two derivatives
in tachyon field, which provides the important tool in verifying Sen's conjectures; more
detailed discussion and references 
can be found in 
\cite{15}, \cite{16}, \cite{17}, \cite{18} for the bosonic case and
\cite{19} for superstring.\footnote{The same tachyon
lagrangian, which we will present below for bosonic string and its analog for
supersymmetric case was proposed
in \cite{20} as a toy model that mimics the properties of tachyon condensation. Very impressive
progress has been achieved in verifying Sen's conjectures in the cubic string field theory of
\cite{1}; see contributions of A. Sen, B. Zwiebach and W. Taylor in the proceedings of this,
Strings 2001, conference. One should note that the world-sheet approach to the problem was
discussed previously in \cite{21}.} The important questions related to the 
description of multiple-branes in the formalizm of 
background independent open string field theory and unified 
treatment of {\bf RR} couplings is studied in \cite{22}.

Following \cite{5} one starts with world-sheet description of critical bosonic string
theory on disk. Consider the map of the disk $D$
to space-time $M$:
\begin{equation}
\begin{split}
X(z, \bar z): \quad \quad \quad D \rightarrow M
\end{split}
\end{equation}
In general one can consider any critical 2d
CFT coupled to 2d gravity on the disk
but it is interesting to study the particular case of
critical bosonic string with flat
26 dimensional space-time $M = R^{1,25}$.

Two-dimensional quantum field theory on the string world-sheet is
given by the path integral:
\begin{equation}
\begin{split}
<...> = \int [dX][db][dc] \quad e^{-I_0(X,b,c)} ...
\end{split}
\end{equation}
\begin{equation}
\begin{split}
I_0= \int_D {\sqrt g} [g^{\alpha \beta}
\partial_{\alpha}X^{\mu}\partial_{\beta}X_{\mu}+
b^{\alpha \beta}D_{\alpha}c_{\beta}]
\end{split}
\end{equation}
 Define BRST operator through
the current $J_{BRST}$ and contour $C$ (note this is a closed
string BRST operator):
\begin{equation}
\begin{split}
Q_{BRST} = \int_C J_{BRST}; \quad \quad Q_{BRST}^2=0
\end{split}
\end{equation}
Denote the limit when contour $C$ approaches the boundary
${\partial}D$
by $Q$:
$Q=\int_{C \rightarrow {\partial D}} J_{BRST}$.
Now we consider the local operator ${\mathcal  V}(X,b,c)$ of the form
\begin{equation}
{\mathcal  V} = b_{-1} {\mathcal  O}(X,b,c),\quad \quad b_{-1}=
\int_{C \rightarrow \partial D} v^ib_{ij}{\epsilon^j_k}dx^k
\end{equation}
and deform the world-sheet action:
\begin{equation}
\begin{split}
I_{ws} = I_0 + \int_{C \rightarrow {\partial D}} {\mathcal V}(X(\sigma))
\end{split}
\end{equation}
\begin{equation}
\begin{split}
<...>=\int [dX][db][dc]e^{-I_{ws}}...
\end{split}
\end{equation}
The simplest case is when ghosts decouple: ${\mathcal O} = c V(X)$.
The boundary term in the action modifies the boundary condition
on the map $X^{\mu}(z, \bar z)$ from the Neumann boundary condition
(this follows from $I_0$)
$\partial_r X^{\mu}(\sigma) = 0$
to ``arbitrary'' non-linear condition:
\begin{equation}
\begin{split}
\partial_r X^{\mu}(\sigma) =
{\partial \over {\partial X^{\mu}(\sigma)}}
\int_{\partial D} V(X)
\end{split}
\end{equation}
$I_{ws}$ defines the family of boundary 2d quantum field theories on the disk. The action
$S({\mathcal O})$ is defined on this space (more precisely - on the space of 
${\mathcal O}$'s) and is
formally independent of the choice of particular open string background:
\begin{equation}
\begin{split}
\label{basic}
dS = <d \int_{\partial D}
{\mathcal O} \quad \{Q, \quad \int_{\partial D} {\mathcal O}\}>
\end{split}
\end{equation}
Since $d{\mathcal O}$ is arbitrary all solutions of the equation $dS=0$ correspond to conformal,
exactly marginal boundary deformations with $\{Q, {\mathcal O}\}=0$, and thus to valid string
backgrounds.

A very important question at this stage is to understand {\it what is the space of deformations
given by $V(X(\sigma))$}. An obvious assumption (which is also a very strong restriction, see the
comment at the end of this talk) is - $V$ can be expanded into ``Taylor series'' in the
derivatives of $X(\sigma)$:
\begin{equation}
\begin{split}
\label{ass}
V(X) = T(X(\sigma)) + & A_{\mu}(X(\sigma))\partial X^{\mu}(\sigma) +
C_{\mu \nu}(X(\sigma))
 \partial X^{\mu}(\sigma)
\partial X^{\nu}(\sigma) +\cr
&
D_{\mu}(X(\sigma)) \partial^2 X^{\mu}(\sigma) +...
\end{split}
\end{equation}
Thus the action now becomes the functional of coefficients: $S = S(T(X(\sigma)),
A_{\mu}(X(\sigma)), ...)$. It is almost obvious that the above assumption singles out the open
string degrees of freedom from very large space of functionals of the map $X^{\mu}(\sigma)$ -
$\partial D \rightarrow M$.
The goal is to write $S$ as an integral over the space-time $X$ (constant mode of $X(\sigma)$:
$X(\sigma) = X + \phi(\sigma)$, $\int \phi(\sigma)=0$) of some ``local'' functional of fields
$T(X), A(X), ...$ and their derivatives.

In a more general setup one can introduce some coordinate system $\{t^1, t^2, ...\}$
in the space of the boundary operators -
${\mathcal O} ={\mathcal O}(t, X(\sigma)); V=V(t,X(\sigma))$:
$$
d{\mathcal O} = \sum dt^i {\partial \over {\partial t^i}}{\mathcal O}(t),  \quad \quad
dV(t,X(\sigma))=\sum dt^i {\partial \over {\partial t^i}} V(t, X(\sigma)
$$
At the origin, $t^i=0$, we
have an un-deformed theory and the linear term in the deformation
is given by an operator $\int_{\partial D} V_i$ of dimension
$\Delta_i$ from the spectrum:
\begin{equation}
\begin{split}
\label{defone}
I_{ws} = I_0 + t^i \int_{\partial D} V_i + O(t^2) =
I_0 + t^i {\partial \over {\partial t^i}} \int_{\partial D} V(t)_{|_{t=0}} + O(t^2)
\end{split}
\end{equation}

For the general boundary term $\int_{\partial D} V$ one might worry that
the two-dimensional theory on the disk is not renormalizable;
it makes sense as a cutoff theory, but it turns out that if one perturbes by
a complete set of
operators from the spectrum the
field theory action is still well-defined (see discussion at the
end  \cite{8}; for the tachyon $T$ and gauge field $A$ world-sheet theory is obviously
renormalizable). It has been proven that 
the action (\ref{basic}) can written in terms of world-sheet $\beta$-function and
partition function \cite{6}, \cite{8}:
\begin{equation}
\begin{split}
\label{main}
S(t) =
 -\beta^i(t) {\partial \over {\partial{t^i}}} Z(t) + Z(t)
 \end{split}
 \end{equation}
here $\beta^i$ is the beta function for coupling $t^i$,
a vector field in the space of boundary theories, and
$Z(t)$ - the matter partition function on the disk.

Since equations of motion $dS=0$
coincide with the condition that deformed 2d theory
is exactly conformal we have
\begin{equation}
\begin{split}
\label{imp}
{\partial \over {\partial {t^i}}} S(t) =
G_{ij}(t) \beta^j(t)
\end{split}
\end{equation}
with some non-degenerate Zamolodchikov metric $G_{ij}(t)$
(the equation $\partial S(t) = 0$ shall be equivalent to
$\beta^i=0$). In addition we see that on-shell
$$S_{on-shell}(t)=Z(t)$$
and as a result classical action on solutions of equation motion will generate correct
tree level string amplitudes.

It is important to note that in general the total derivatives don't decouple inside the
correlation function and we have coupling dependent BRST operator $Q(t)$.
In fact the same contact terms contributes to
$\beta$-function. More precisely one can fix the contact terms from the condition that
the definition (\ref{basic}) is self-consistent after contact terms are included:
\begin{equation}
\begin{split}
\label{myold}
Q = Q(t);  \quad  < .... \int \partial_{\theta} ...>
\neq 0
\end{split}
\end{equation}
\begin{equation}
\begin{split}
\label{correct}
d \quad {\mathcal [}\quad  dS =  <d \int_{\partial D}
{\mathcal O} \quad \{Q, \quad \int_{\partial D} {\mathcal O}\}>
\quad ] =0
\end{split}
\end{equation}
If one assumes that $Q$ is constant and total derivatives decouple - (\ref{correct})
is a simple Ward identity; otherwise it is a condition which relates the choice
of contact terms with $t$ dependence of corresponding modes of $Q$ \cite{7}, 
\cite{8}. Usually, in quantum field
theory one has to choose the contact terms (regularization) based on some (symmetry) principle
(an example from recent years is the Donaldson theory which rewritten in terms of
Seiberg-Witten IR description requires the choice of the contact 
terms based on Seiberg-Witten modular
invariance together with topological $Q$ symmetry \cite{23} and dependence of $Q$ on moduli
is fixed from this consistency principle); here we have (\ref{imp}), (\ref{correct}) as guiding
principle.

The principle (\ref{correct}) leads to the formula (\ref{main}) for the action with
all non-linear terms included: $\beta^i = (\Delta_i-1)t^i + c^i_{jk}t^jt^k + ...$
and in addition guarantees that Zamolodchikov metric in (\ref{imp}) is non-degenerate.
In second order all terms except those that satisfy the resonant 
condition $\Delta_j - \Delta_i+ \Delta_k=1$
can be removed by redefinition of couplings; obviously these correspond to
logarithmic divergences and thus if the theory is perturbed with the 
complete set of operators - it is logarithmic. More generally - from Poincare-Dulac
theorem (which of course is for the finite-dimensional case but we will assume it is correct
for infinite-dimensional space also) one can
show that all coefficients are cutoff-independent after an appropriate choice 
of coordinates is made.
Let us repeat in regard to the choice of coordinates - as it follows
from the above discussion, any choice of coordinates is good as long as
{\it  equations of
motion lead to
$\beta$-function equations with invertible Zamolodchikov metric
$G_{ij}$.}

First we turn on only tachyon: $V(X(\sigma)) = T(X(\sigma))$.
It is not difficult to find the action $S(T)$ as an expansion in derivatives;
for example - exact in $T$ and second
order in derivatives $\partial$.
We know the derivative expansion of $\beta$ and $Z$:
\begin{equation}
\begin{split}
\label{expan}
\beta^T(X) = [2 \Delta T + T] + a_0(T) +
a_1(T)(\partial T) +
a_2(T)(\partial^2 T) + a_3(T) (\partial T)^2 + ...
\end{split}
\end{equation}
\begin{equation}
\begin{split}
\label{parti}
Z(T) = \int dX e^{-T}(1 + b(T)(\partial T)^2 + ...)
\end{split}
\end{equation}
with appropriate conditions for unknown coefficients dictated by the
perturbative expansion.
In this concrete case the basic relation becomes:
\begin{equation}
\begin{split}
\label{basictwo}
S(T) =  - \int dX \beta^T (X)
{\partial \over {\partial T(X)}} Z(T)  +   Z(T)
\end{split}
\end{equation}
The condition for $\beta$ in (\ref{expan}) to be an
equation of
motion for $S(T)$ (\ref{basictwo}) with $Z$ given by (\ref{parti})
in lowest order in $T$
(around $T=0 \rightarrow 2\Delta T + T=0$)
fixes the two derivative action (and relevant unknowns
$a_i(T), b(T)$):
\begin{equation}
\begin{split}
\label{gersam}
S(T) = \int dX [e^{-T} (\partial T)^2 + e^{-T}(1+T)]
\end{split}
\end{equation}
with equations of motion, $\beta^T$ and metric $G=e^{-T}$ in (\ref{imp}):
\begin{equation}
\begin{split}
\label{eom}
\partial_T S = e^{-T}{\beta^T} =
e^{-T} ({2\Delta T + T - (\partial T)^2})=0
\end{split}
\end{equation}
\begin{equation}
\begin{split}
\label{eomone}
\beta^T=2\Delta T + T - (\partial T)^2;
\quad Z(T)=\int e^{-T}
\end{split}
\end{equation}
This answer was deduced from the consistency condition
for the expansion (\ref{expan}), (\ref{parti}) and basic relation
(\ref{basictwo}), but one can compute it directly from
world-sheet definition, practically repeating the computations
(in this particular case)
leading to general relation (\ref{main}). We first write the
world-sheet path integral only in terms of boundary data
$\int dX(\theta) e^{-I_{ws}}=\int dX d\phi(\theta)e^{-I_{ws}}$
 (we use the notation $
H(\theta , \theta') =
{1\over 2} \sum_{k} e^{i k (\theta - \theta')} |k|
$; bulk part decouples since arbitrary map $X(z, \bar z)$ can be written as the
sum of two terms: one has zero value on the boundary, another coincides with $X(\sigma)$
on boundary and is harmonic in the bulk):
\begin{equation}
\begin{split}
\label{world}
X(\theta) = X + \phi(\theta); \quad \quad \quad
\int \phi (\theta) = 0;
\end{split}
\end{equation}
\begin{equation}
\begin{split}
\label{act}&
I_{ws}= \int \int d\theta d \theta'
X^{\mu}(\theta) H(\theta, \theta') X_{\mu}(\theta')+\int d\theta
T(X(\theta)) \cr
&=
T(X) + \int \phi^{\mu}(\theta)
[H(\theta-\theta')\delta_{\mu \nu}+
\delta(\theta-\theta')
\partial_{\mu}\partial_{\nu} T(X)]\phi^{\nu} (\theta') +\cr
& \quad \quad \quad \quad
 + O(\partial^3 T(X))
 \end{split}
 \end{equation}
In the two derivative approximation contribution comes only from:
\footnote{If we turn on other fields from (\ref{ass}) it immediately follows from
corresponding expression of the type (\ref{act}) that only tachyon will condense;
see discussion in \cite{16}.}
\begin{equation}
\begin{split}
\label{w}
Z(T) = \int dX e^{-T(X)} det' [ H +
\partial^2 T]^{-{1 \over 2}}
\end{split}
\end{equation}
This can be computed exactly.
For $\partial_{\mu}\partial_{\nu}T=
\delta^{\mu\nu}\partial^2_{\mu}
T$ it is \cite{6} (in some regularization) :
\begin{equation}
\begin{split}
\label{wwr}
Z=\int dX e^{-T(X)} \prod_{\mu}\sqrt{
{\partial_{\mu}^2 T(X)}}e^{ \gamma
{\partial_{\mu}^2 T(X)}} \Gamma(
{\partial_{\mu}^2T(X)})
\end{split}
\end{equation}
and in the two-derivative approximation this gives \footnote{In fact (\ref{wwr})
can be used only for two-derivative approximation for the action
since the contribution of
the last term in (\ref{act}) will mix with higher order terms
coming from $\Gamma$ function in (\ref{wwr}) after integration by parts
due to the presence of universal exponential $e^{-T}$
factor in the action; this is very similar to what happens
in the attempts to write non-abelian version of Born-Infeld action.} :
\begin{equation}
\begin{split}
\label{twod}
Z(T) =& \int dX e^{-T(X)}(1+b(T))(\partial T)^2) \cr
&
\beta^T = 2\Delta T + T \cr
&
b(T) = 0, \quad a_0(T)=a_1(T)=a_2(T)=a_3(T)=0
\end{split}
\end{equation}
Thus the action (\ref{main}) is:
\begin{equation}
\begin{split}
\label{acwit}
S(T) = \int dX [e^{-T}{2}(\partial T)^2 +
e^{-T}(1+T)]
\end{split}
\end{equation}
with equations of motion:
\begin{equation}
\begin{split}
\label{eomwi}
e^{-T}(T+{4}\Delta T-{2}(\partial T)^2)=0
\end{split}
\end{equation}
Now we see that Zamolodchikov metric $G$ becomes :
\begin{equation}
\begin{split}
\label{metricnew}
G(\delta_1 T,\delta_2 T)=\int dX
 e^{-T}(\delta_1T\delta_2 T-
 2(\partial_{\mu}\delta_1 T)(\partial_{\mu}\delta_2 T))
 \end{split}
 \end{equation}
and equations of motion (\ref{eomwi}) can be written in
this approximation as
$G\beta=0$:
\begin{equation}
\begin{split}
\label{newequati}
e^{-T}(1+2\Delta -
 2\partial_{\mu}T\partial_{\mu}+...)({2\Delta T + T})=0
 \end{split}
 \end{equation}
The linear form of $\beta$ for arbitrary $T(X)$ looks strange
since we miss a possible higher order in $T$
(but second order in $\partial T$)
terms which shall come from a 3-point function.
In addition this metric is rather complicated and
is not an obvious expansion
of some invertible metric in the space of fields.
Thus, according the principle for the choice of coordinates,
we need to choose new
coordinates such that
the metric is invertible. Such new variables are given by:
\begin{equation}
\begin{split}
\label{change}
T\rightarrow T-\partial^2 T +(\partial
 T)^2
\end{split}
 \end{equation}
This modifies the $\beta$ function (without changing its linear part)
and metric to (\ref{eom}), (\ref{eomone}); in new coordinates action is given by:
$$S(T) = \int dX [e^{-T} (\partial T)^2 +
 e^{-T}(1+T)]$$

The potential in this action has unstable extremum
at $T=0$ (tachyon)
and stable at $T=\infty$. The difference between the values of this
potential is $1$, exactly as predicted by A. Sen in {\bf Conjecture 1}.

In a new variable with the canonical kinetic term
$\Phi = e^{-{T \over 2}}$:
\begin{equation}
\begin{split}
\label{phiaction}
S(\phi) = \int [4(\partial \Phi)^2 -
\Phi^2 log {\Phi^2 \over e}]
\end{split}
\end{equation}
(interestingly this is also an exact action, see \cite{15},
for a $p$-adic string for
$p \rightarrow 1$).
In the unstable vacuum $T=0, \Phi=1; m^2=-{1 \over 2}$; in new
vacuum - $T=\infty, \Phi=0$ and one could naively
think that $m^2=+\infty$, but since the action
is non-analytic at this point the meaning is unclear.

$DN$ boundary conditions, $p\leq25$:
\begin{equation}
\begin{split}
\label{dn}
\partial_r X^a(\sigma)=0, \quad a =1,...,p; \quad \quad 
X^i(\sigma)=0, \quad i=p+1,...,26
\end{split}
\end{equation}
are obviously conformal. We conclude that they give
critical points of string field theory action. In addition
since the value of classical action is always
$S(t_c)=Z(t_c)$ -  we conclude that these solitons are
in fact $Dp$-branes; e. g. one can take \cite{5}:
\begin{equation}
\begin{split}
\label{dnone}
T(X) = a + u_{\mu} (X^{\mu})^2
\Rightarrow \partial_r X^{\mu}=u_{\mu}X^{\mu}; \quad
u_i \rightarrow \infty, \quad u_a \rightarrow 0
\end{split}
\end{equation}
This verifies the {\bf conjecture 2}.

{\bf Conjecture 3} is in the heart of the discussion we
had in the beginning, and is also most difficult one. In order
to address it we add the gauge field.
The action in two derivative approximation can
be constructed using the same basic relation (\ref{main}), \cite{18}.
Here we will also introduce the (background) closed string fields
$G$ and $B$ (for covariance):
\begin{equation}
\begin{split}
\label{gauge}
&\quad \quad \quad \quad
S(G,B,A,T)=\cr
&
S_{closed}(G,B)+\int d^{26}X \sqrt{G}
[e^{-T}(1+T)+e^{-T}||dT||^2+
{1\over 4}e^{-T}||B-dA||^2+\cdots ]
\end{split}
\end{equation}
One can choose a different regularization and obtain the action
which is an expansion of {\bf BI} action
$$\int V(T)\sqrt {det(G - B + dA)}$$
but it would lead to a complicated and not
obviously non-degenerate
metric, exactly like in purely tachyon case discussed above \ref{newequati})).

In $\Phi=e^{-{T \over 2}}$ coordinates:
\begin{equation}
\begin{split}
\label{gaugetwo}&
S(G,B,A,\Phi)=\cr
&
S_{closed}+
\int d^{26}X \sqrt{G}
 [\Phi^2(1-2\log \Phi)+4||d\Phi||^2+
 +{1\over
4}\Phi^2||B-dA||^2+\cdots]
\end{split}
\end{equation}
The latter immediately suggests the analogy with
an abelian Higgs model for complex scalar
and gauge field in angular coordinates
$ H e^{i\phi}, {\mathcal A}$:
\begin{equation}
\begin{split}
\label{abelian}
S(H,
\phi,{\mathcal A})=S_{YM}({\mathcal A})+\int [\lambda
(H^2-H_0^2)^2)+|dH|^2+
 + H^2|{\mathcal A}-d\phi|^2]
 \end{split}
 \end{equation}
with identifications:
\begin{equation}
\begin{split}
\label{ident}
B \rightarrow {\mathcal A}, \quad \quad
A \rightarrow \phi, \quad \quad \Phi=
e^{-{1 \over 2}T} \rightarrow H
\end{split}
\end{equation}
Exactly like in a symmetric point for an abelian Higgs model where
phase $\phi$ is not a good coordinate - in string theory
the gauge field $A$ becomes ill-defined in new vacuum,
 $\Phi=0, T=\infty$; the same is
true for all open string modes \cite{24},
\cite{18}; all open string degrees of freedom, except the zero mode
of tachyon, $T_0$, are angular variables
with $T_0$ being only the radial one. In addition, as it has been explained in \cite{18},
using the Zwiebach's open/closed string field theory, in the symmetric vacuum
when the closed string gauge symmetries are restored, all
open string degrees of freedom are gauge parameters for closed string
gauge transformation and thus disappear from the spectrum. Thus, as long as we
know that new vacuum is invariant under closed string gauge transformation
we can safely conclude that there are no open string degrees of freedom.

 We shall note that in an abelian Higgs model instead of assuming that background gauge field is
a dynamical variable
one can also choose the cartesian coordinates, but in the stringy case there is
no possibility to do so since the ``phase" (gauge field) and absolute value ($\Phi=e^{-{T \over
2}}$) carry different space-time spin, thus we can only conclude that
the theory is smooth and consistent with closed string modes becoming dynamical in new vacuum.
This should be compared to the restriction made for the boundary functional
$V(X(\sigma))$ in (\ref{ass}) - we see now that our assumption
was too restrictive and one shall consider in addition
truly non-local functionals which most likely shall correspond
to dynamical closed string modes. At the same time one can introduce non-local
string field theory wave-function:
\begin{equation}
\begin{split}
\label{wave}
\Psi(X(\sigma)) = e^{-\int_C T(X(\sigma)) + A_{\mu}(X(\sigma)) \partial X^{\mu}(X(\sigma)) +...}
\end{split}
\end{equation}
which can be considered as a formal analog of complex scalar (cartesian coordinates)
field and 2-form field
$B$ gives natural connection on the space of such functionals
(this $\Psi$ now depends on the choice of contour $C$ in space-time; the creation and 
annihilation operators
of $A$ do not make sense anymore and only loop operator $\Psi$ describes the 
physical degree of freedom). It seems to be closely
related to string field that enters in cubic string field theory \cite{1}: consider 
wave-function/string field $\Psi(X_*(\sigma))$ given by world-sheet
path integral for disk and divide it in two equal parts with the first half carrying the fixed
boundary conditions $X(z, \bar z)_{\partial D} = X_*(\sigma)$ and on the other half the operator
(\ref{wave}) inserted. One shall think about this path integral
as a very non-linear and non-local change of variables from
$V(X(\sigma))$  to $\Psi$ of cubic string field theory:
\begin{equation}
\begin{split}
\label{change}
\Psi_{V(X)}(X_*(\sigma)) = \int_{X_*(\sigma)}
[dX] \quad  e^{-I_{0}(X)-\int_0^{\pi} V(X(\sigma))}
\end{split}
\end{equation}
For tachyon zero mode it leads to $e^{-{T \over 2}} = 1+T_0^{cubic}$.
One can map the disk to ``1/3 of pizza" (see Fig. 1) and glue three such wave-functions
in order to get a cubic term in cubic {\bf CS} string field theory which becomes the disk
partition function with the operator (\ref{wave}) inserted everywhere on the boundary of the 
disk (the second term
in the action (\ref{main})); as far as first term with $\beta$-function it should be possible
to obtain from cubic string field theory action kinetic term $(\Psi, Q \Psi)$ via the conformal
map of the disk to the half of the disk (note this
$Q$ is now the open string BRST operator as opposed to the one which enters in the definition of
background independent string field theory action). This proposal 
is incomplete and needs serious
investigation; unfortunately a more direct relation is
difficult to demonstrate since we considered the restrictive situation when ghosts decouple
from matter in the boundary perturbation. \begin{center}
\begin{figure}
\setlength{\unitlength}{.4mm}
  \thicklines
\begin{picture}(30,30)(10,-13)
    \put(0,0){\circle{50}}
    \path(0,-25)(0,25)
    \put(62.5,0){\circle{50}}
    \path(62.5,0)(62.5,-25)
    \path(62.5,0)(40.849,12.5)
    \path(62.5,0)(84.151,12.5)
  \end{picture}
\caption{picture 1}
\end{figure}
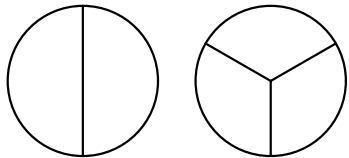
\end{center} One can wonder what is the description of the above space-time picture 
from the point of view of
world-sheet theory on disk. Since condensation takes place for infinite value of tachyon
field, $T \rightarrow \infty$, we see that the typical size of the boundary which will contribute
to world-sheet path integral has to be small and thus at the end-point of condensation the
boundary of the disk shrinks to the zero size (see the Fig. 2). Thus there is no boundary
anymore and the operator insertion at the boundary in fact becomes the insertion at the points
on the bulk. More careful treatment shows that these operators, which by definition
were the operators defined in the bulk and taken to the boundary, are integrated over the whole
sphere (with Liouville factor properly included)
and thus deformation on the boundary becomes deformation on the sphere.
It is tempting to think
that this is in fact the correct world-sheet way of thinking about the end-point of condensation
and string field theory action for open strings
(\ref{main}) becomes
the action for closed string degrees of freedom.
\begin{center}
\begin{figure}
  \setlength{\unitlength}{1mm}
  \thinlines
  \begin{picture}(100,30)(10,0)
    \put(50,0){\epsfig{figure=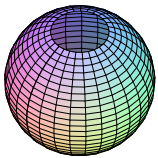,height=20mm}}
  \end{picture}
\caption{picture 2}
\end{figure}
\end{center}
{\bf Acknowledgments.} I would like to thank A. Gerasimov for collaboration during 
the recent studies in \cite{15}, \cite{18}
and Rutgers New High Energy Theory Group for hospitality. This work is supported by OJI award
from DOE.

\bibliographystyle{amsalpha}

\end{document}